\documentclass[final,1p,times,twocolumn]{elsarticle}
\usepackage{amssymb}
\usepackage{amsmath}
\usepackage{bm}

\newcounter{bla}

\journal{Computer Physics Communications}

\usepackage{hyperref}
\usepackage{url}
\usepackage[normalem]{ulem}\usepackage[normalem]{ulem}
\usepackage[usenames,dvipsnames]{color}
\usepackage{fancyvrb}
\VerbatimFootnotes

\newcommand{\editor}[2]{%
  \expandafter\newcommand\csname #1note\endcsname[1]{%
    \textcolor{#2}{(\textbf{#1:} ##1)}}%
  \expandafter\newcommand\csname #1\endcsname[1]{%
    \textcolor{#2}{##1}}%
  \expandafter\newcommand\csname #1cancel\endcsname[1]{%
    \textcolor{#2}{\sout{##1}}}%
  \expandafter\newcommand\csname #1change\endcsname[2]{%
    \textcolor{#2}{\sout{##1} ##2}}%
  \newenvironment{#1text}{\color{#2}}{\color{black}}
}
\definecolor{tangerine}{rgb}{0.944,0.522,0}
\editor{Revision}{red}

\def\SporTran{\texttt{SporTran}}
\newcommand{\angstrom}{\mbox{\normalfont\AA}}

\begin{document}

\begin{frontmatter}

\title{SporTran: a code to estimate transport coefficients from the cepstral analysis of (multivariate) current time series}

\author[a,b]{Loris Ercole}
\author[b]{Riccardo Bertossa}
\author[b,c]{Sebastiano Bisacchi}
\author[b,d]{Stefano Baroni\corref{author}}

\cortext[author] {Corresponding author.\\\textit{E-mail address:} baroni@sissa.it}
\address[a]{Theory and Simulation of Materials (THEOS), and National Centre for Computational Design and Discovery of Novel Materials (MARVEL), \'Ecole Polytechnique F\'ed\'erale de Lausanne, CH-1015 Lausanne, Switzerland.}
\address[b]{SISSA – Scuola Internazionale Superiore di Studi Avanzati, Via Bonomea 265, 34136 Trieste, Italy}
\address[c]{Istituto Statale di Istruzione Superiore Galileo Galilei, via Puccini 22, 34170 Gorizia}
\address[d]{CNR – Istituto Officina dei Materiali, SISSA, 34136 Trieste}

\begin{abstract}
\SporTran\ is a Python utility designed to estimate generic transport coefficients in extended systems, based on the Green-Kubo theory of linear response and the recently introduced \emph{cepstral analysis} of the current time series generated by molecular dynamics simulations. \SporTran\ can be applied to univariate as well as multivariate time series. Cepstral analysis requires minimum discretion from the user, in that it weakly depends on two parameters, one of which is automatically estimated by a statistical model-selection criterion that univocally determines the resulting accuracy. In order to facilitate the optimal refinement of these parameters, \SporTran\ features an easy-to-use graphical user interface. A command-line interface and a Python API, easy to embed in complex data-analysis workflows, are also provided.
\end{abstract}


\end{frontmatter}



\noindent {\bf PROGRAM SUMMARY}

\smallskip
\begin{small}
\noindent {\em Program Title:}  \SporTran  \\
{\em Licensing provisions:}  GPLv3 \\
{\em Programming language:}  Python \\

\noindent {\em Nature of problem:}\\
Given an $M$-variate time series, $\mathcal{J}^j(t)$, $j=0,\dots M-1$, typically describing a number of currents resulting from a molecular-dynamics simulation, \SporTran\  estimates the transport coefficient $\kappa = 1/\left (\Lambda^{-1} \right )^{00}$, where $\Lambda^{ij} = \int_0^\infty\! \left\langle \mathcal{J}^j(t) \mathcal{J}^k(0) \right \rangle dt$ is the matrix of the Onsager linear-response coefficients, and $\langle\cdot\rangle$ indicates an equilibrium average over initial conditions.

\smallskip\noindent{\em Solution method:}
\vspace{-5pt}
\begin{enumerate}[\itshape i)]
\item It is first observed that the Onsager transport coefficients are the zero-frequency values of the cross power spectra of the currents under scrutiny: $\Lambda^{ij}=\frac{1}{2}S^{ij}\bigl (\omega=0 \bigr )$, where $S^{ij}(\omega)= \int_{-\infty}^\infty \!\left\langle \mathcal{J}^i(t) \mathcal{J}^j(0) \right \rangle \mathrm{e}^{i\omega t}dt$.
\item We next define the (cross) \emph{periodogram} as the product of pairs of Fourier transforms of the current time series: $ \mathcal S^{ij}_k = \frac{\epsilon}{N} \tilde{\mathcal J}^i_k \tilde{\mathcal J}^{j*}_k$, where $\epsilon$ is the time step of the time series, $N$ the number of their terms, and ${\tilde{\mathcal{J}}}_k^j = \sum_{n=0}^{N-1} \mathcal{J}^j_n \mathrm{e}^{2\pi i\frac{kn}{N}}$ their discrete Fourier transforms, and $\mathcal{J}^j_n=\mathcal{J}^j(\epsilon n)$. 
\item As the current time series are realisations of a Gaussian process, in the long-time limit and for $k\ne k'$ the $ \mathcal S^{ij}_k $ are uncorrelated  complex Wishart random matrices (a matrix generalization of the $\chi^2$ distribution) whose expectation, according to the Wiener-Khintchine theorem, is the cross power spectrum we are after. It follows that $\left ( \mathcal S^{-1}_k \right )^{00}$ is proportional to a set of uncorrelated $\chi^2$ deviates;
\item A consistent estimator for $\log(\kappa)=-\log \left ( \bigl (\Lambda^{-1} \bigr )^{00} \right )$ is finally obtained by applying a low-pass filter to the process $ \log \left ( \left ( \mathcal S^{-1}_k \right )^{00}\right )$.
\end{enumerate}
 The theoretical background of the methodology implemented in \SporTran\  is thoroughly presented in Refs. [1-3].

\end{small}

\section{Introduction}\label{sec:Introduction}
Estimating transport coefficients from the Green-Kubo (GK) theory of linear response and equilibrium molecular dynamics (EMD) can be a challenging task. Long EMD trajectories and cumbersome data analysis, not always properly performed, are needed to average out the thermal noise affecting such calculations \cite{Ercole2017}. Further difficulties, both conceptual and practical, arise in the multivariate case, where several conserved quantities (say, energy and the number of molecules of different chemical species) are being transported simultaneously, and thus interact with each other \cite{Bertossa2019}. This state of affairs is particularly annoying in a quantum mechanical setting \cite{Marcolongo2016,qeheat,Marcolongo2020,Grasselli2019,Tisi2021}, for the numerical workload of \emph{ab initio} simulations based on electronic-structure theory is so heavy as to practically  limit the length of the EMD trajectories that can be afforded to a few hundred picoseconds, at most. The cepstral data analysis \cite{Bogert1963,Oppenheim:2004} protocol enables one to estimate transport coefficients, along with their statistical errors, without the need of performing cumbersome block analyses and it is statistically much more efficient than any straightforward implementation of the GK integral \cite{Ercole2017,Bertossa2019,Baroni2018}. This procedure substantially reduces the trajectory length necessary to achieve a target accuracy and, most importantly, allows one to evaluate the latter in a statistically rigorous and practically manageable way. Remarkably, cepstral analysis also applies with minimal adaptations and little, if any, numerical and statistical overhead to the multivariate case \cite{Bertossa2019}.

\SporTran\ is designed to perform a complete cepstral analysis  of a possibly multivariate current time series generated by EMD, without the need of implementing, or even fully mastering, all the details of the protocol. Moreover, an easy-to-use and multi-platform Graphical User Interface (GUI) is provided, allowing the inexperienced user to input all the needed parameters and visualise all the intermediate results in an interactive and easy-to-understand way. In addition, the package is equipped with a command-line interface, permitting to run the program on headless computers or easily embed it in scripts. Finally, a flexible Python API is provided, giving more advanced users complete control over all the parameters and intermediate steps of the protocol.

The \SporTran\ package requires at least Python 3 and the numerical library \texttt{numpy} \cite{harris2020array} for its core functionalities. In addition to this, the complete package requires \texttt{scipy}, \texttt{matplotlib}, \texttt{markdown2} and \texttt{pillow}. It can be installed using the \texttt{pip} utility (\Verb.pip install sportran.) or by cloning the GitHub repository \cite{sportran}.

\section {Theoretical background}\label{sec:Theory}

The macroscopic behaviour of a material, including its response to external perturbations, dissipation, and the approach to equilibrium, is determined by the time evolution of its hydrodynamical variables, \emph{i.e.} of the long-wavelength components of the densities and current densities of its conserved quantities (such as mass, chemical composition, charge, energy, and momentum), which we call for short \emph{conserved densities} and \emph{conserved currents}.

Let $J^i$ be the macroscopic average of the $i$-th conserved current, $J^i = \frac{1}{V} \int\! j^i(\mathbf{r}) d\mathbf{r}$, where $V$ is the system's volume, which from now on we dub as a (conserved) \emph{flux}. In order to unclutter the notation, we will dispose of the Cartesian indices of currents and other vectors, or, if one prefers, we will merge them into the suffixes marking currents off from one another. In the case of heat transport in an $M$-component fluid, the relevant conserved fluxes are those of the energy and of the mass/number of each one of its molecular components. As the total-mass flux is the total momentum, which is also a constant of motion, the number of relevant conserved fluxes is reduced from $M+1$ to $M$: energy, which we label as the zero-th, and $M-1$ \emph{convective} (mass) fluxes. 

In the linear regime, conserved fluxes are related to the thermodynamic forces, $F^j$ (\emph{i.e.} to the gradients of the intensive variables conjugate to the conserved quantities being transported), by the Onsager relations \cite{Onsager1931a,Onsager1931b}:
\begin{equation}
    J^i = \sum_{j=0}^{M-1} \Lambda^{ij} F^j. \label{eq:Onsager}
\end{equation}
For an $M$-component fluid the relevant thermodynamic forces are the gradients of the inverse temperature and of negative of the ratio between the (electro-) chemical potential of each molecular species and temperature. The GK theory of linear response \cite{Green1952,Green1954,Kubo1957a,Kubo1957b,Baroni2018} states that the $\Lambda$ matrix in Eq.~\eqref{eq:Onsager} can be expressed in terms of the time correlation functions of the various \textit{flux processes}, $\mathcal{J}^i(t)$, as:
\begin{equation}
\Lambda^{ij} = \frac{V}{k_B} \int_0^\infty\! \left\langle \mathcal{J}^i(t) \,\mathcal{J}^j(0) \right \rangle dt, \label{eq:L_def}
\end{equation}
where $k_B$ is the Boltzmann constant and $\langle\cdot\rangle$ indicates an equilibrium average over the initial conditions of a molecular trajectory \cite{Baroni2018}. From now on, a calligraphic letter, such as $\mathcal A(t)$ will indicate the value that a phase-space variable, $A(q,p)$, assumes at time $t$: $\mathcal A(t)=A(q_t,p_t)$. When sampled by EMD, $\mathcal A(t)$ is represented by a discrete time series. The integrals of Eq.~\eqref{eq:L_def} can be evaluated from these time series, but the numerical estimate of their values and of the associated statistical errors is ill-conditioned, particularly, but not exclusively, in the multivariate case.

In a multi-component system, a transport coefficient is defined as the ratio between a conserved flux (say, the energy flux, in the case of heat transport) and the corresponding thermodynamic force, \emph{when all the other conserved fluxes} (the convective ones, in this case) \emph{vanish}:
\begin{equation}
    J^i = \begin{cases} ~\kappa \, F^0 & \mbox{if } i =0\phantom{.} \quad\quad (a)\\
~ 0 & \mbox{if } i \ne 0. \quad\quad (b) \end{cases} \label{eq:multi-kappa1}
\end{equation}
Imposing the relation (\ref{eq:multi-kappa1}b) in Eq. \eqref{eq:Onsager}, one obtains:
\begin{align}
    \kappa = \frac{1}{\left(\Lambda^{-1}\right)^{00}}. \label{eq:multi-kappa2}
\end{align}

The Onsager transport coefficients are best evaluated as the zero-frequency values of the cross power spectra of the fluxes under scrutiny: $\Lambda^{ij}=\frac{1}{2}S^{ij}\bigl (\omega=0 \bigr )$, where $S^{ij}(\omega)= \int_{-\infty}^\infty\! \left\langle \mathcal{J}^i(t) \mathcal{J}^j(0) \right \rangle \mathrm{e}^{i\omega t}dt$. The reason why this is convenient is because, if the current cross power spectra are smooth enough in the $\omega\to 0$ limit, their estimates at finite frequency can be leveraged to reduce the error at $\omega = 0$ in the spirit of a generalized central-limit theorem, as explained in Ref.~\citenum{Ercole2017} for the univariate case and generalized to the multivariate one in Ref.~\citenum{Bertossa2019}. The way this is achieved is technically rather complex and we refer the reader to Refs.~\citenum{Ercole2017} and \citenum{Bertossa2019} for all the necessary conceptual and practical details.

\section{Workflow}

In a nutshell, the cepstral method embedded in \SporTran\  requires the following sequence of operations:
\begin{enumerate}
    \item Read $\ell$ independent samples of the $M$-dimensional Gaussian process that one wants to analyse $\left\{ ^{p\!}\mathcal{J}^i_n \right\} ~(p=1,\dots \ell; ~i=0,\dots M-1; ~n=0,\dots N-1)$. $M$ is the number of conserved fluxes that characterise the transport process; in practice, $\ell$ may be the number of statistically independent segments in which a long EMD trajectory has been partitioned, times the number of Cartesian components of the fluxes that are equivalent by symmetry; finally, $N$ is the number of time steps in each EMD segment. Such an $M \times N \times \ell$ array is stored, together with the time step $\epsilon$ and a conversion factor $\kappa_{\rm{scale}}$ necessary to accommodate for units of the various currents, in a suitably defined \texttt{Current} Python object. 
    Several commonly used units are already implemented, as well as various types of transport coefficients (heat and charge conductivity, viscosity).
    \item Compute the discrete Fourier transforms of these fluxes:
    \begin{equation}
        ^{p\!}{\tilde{\mathcal{J}}}_k^j = \sum_{n=0}^{N-1} {}^{p\!}\mathcal{J}^j_n \,\mathrm{e}^{2\pi i\frac{kn}{N}},
    \end{equation}
    the so-called \emph{cross-periodogram},
    \begin{equation}
        \mathcal{S}_k^{ij}= \frac{\epsilon}{N\ell } \sum_{p=1}^{\ell} \left({}^{p\!}\tilde{\mathcal{J}}_k^i\right)^* {}^{p\!}\tilde{\mathcal{J}}_k^j, \label{eq:X-periodogram}
    \end{equation}
    and the quantity:
    \begin{equation}
        \bar{\mathcal{S}}_k^0 = \frac{\ell}{\ell - M + 1} \frac{1}{\left( (\mathcal{S}_k)^{-1} \right)^{00}},
    \end{equation}
    where $k=1,\ldots N$ is the frequency index. The $\mathcal{S}_k$ $M\times M$ matrices are statistically independent from one-another and must be inverted for each $k$. This is implemented in the \Verb.compute_psd. method of the Python class \texttt{Current}. The expectation value of the 0-th frequency component of $\bar{\mathcal{S}}^0_k$ is (proportional to) the transport coefficient we are after. In the univariate case, this reduces to the periodogram of the original flux. At this point we apply the cepstral analysis to this array, \emph{i.e.} a low-pass filter to its logarithm.
    \item Determine a cutoff (Nyqvist) frequency for analysing the log-spectrum, $f^*$. The statistical accuracy of the cepstral method increases with the number of frequencies in the spectral range being analysed, and decreases with the number of inverse Fourier (cepstral) coefficients necessary to describe the log-spectrum in this range. The choice of $f^*$ is aimed at optimizing the balance between these two contrasting requirements. The code automatically resamples the time series with a time step $\epsilon^\ast\approx \frac{1}{2f^\ast}$, resulting in a smaller number of terms in the time series, $N^\ast$. As a rule of thumb, in most of the practical cases $f^*$ should include the lowest-frequency feature of the periodogram \cite{Ercole2016}.
    \item Calculate $\log \bar{\mathcal{S}}^0_k$.
    \item Compute the \emph{cepstral coefficients} as the inverse discrete Fourier coefficients of the log-spectrum thus calculated:
    \begin{equation}
        \mathcal{C}_n =\frac{1}{N^\ast} \sum_{k=0}^{N^\ast-1}\log \left(
        \bar{\mathcal{S}}^0_k \right) \mathrm{e}^{-2\pi i\frac{kn}{N^\ast}}.
    \end{equation}
    \item Determine the optimal number of cepstral coefficients to be retained in the above summation, $P^*$, using the Akaike information criterion \cite{Akaike1973,Akaike1974,Ercole2017}. If needed, in particular cases it is possible to increase this value manually in order to reduce the bias of the estimator.
    \item Evaluate the transport coefficient and its statistical uncertainty estimated as:
    \begin{equation}
        \begin{aligned}
            \kappa &= \frac{\kappa_{\mathrm{scale}}}{2} \mathrm{exp}\left [ \mathcal{C}_0 + 2\sum_{n=1}^{P^*-1}\mathcal{C}_{n} -L_0\right ], \\
            \frac{\Delta\kappa}{\kappa} &= \sigma_0 \sqrt{\frac{4P^*-2}{N^\ast}},
        \end{aligned} \label{eq:nutshell}
    \end{equation}
    where $L_0= \psi(\ell-M+1) -\log(\ell-M+1) $ and $\sigma_0^2= \psi'(\ell-M+1)$, $\psi$ and $\psi'$ being the di- and tri-gamma functions \cite{PolyGamma}, respectively.
\end{enumerate}
The relative error in the conductivity results from the estimate of the absolute error in its logarithm. $P^*$ depends in general on the Nyqvist frequency, $f^\ast$, used to analyse the spectrum, while the final value of $\kappa$ and its statistical uncertainty are to a large extent independent of it \cite{Ercole2017,Baroni2018}. In any event, the GUI described below assists the user in the determination of these two parameters.

\section{Code description}
\subsection{Graphical User Interface}
\begin{figure}[tb]
    \centering
    \includegraphics[width=0.75\linewidth]{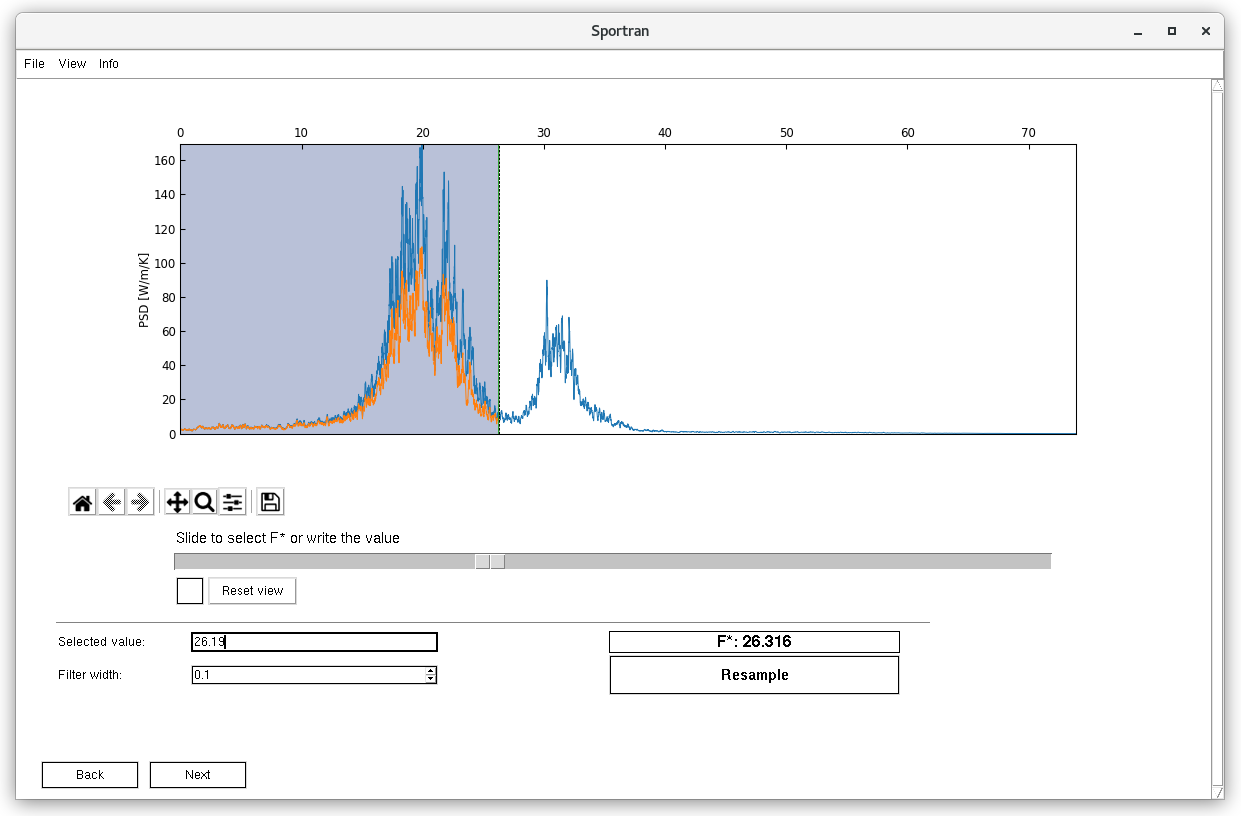}
    \caption{Selection of the Nyqvist frequency, $f^*$. The figure shows the power spectrum of the energy current of a silica glass sample provided in the examples.}
    \label{fig:GUI_fstar}
\end{figure}

The GUI is designed to smoothly drive the user from the raw data to the final result. The input can be a simple column-formatted text file, with a text header for every column, or a NumPy/pickle binary file containing a dictionary. The user selects the file format and is then guided through a few simple steps to load the file, select the desired currents, set the physical constants, and determine the value of the Nyqvist frequency, $f^*$, as well as the optimal number of cepstral coefficients, $P^\ast$. The latter step is actually performed automatically through the Akaike information criterion \cite{Akaike1973,Akaike1974}, whose suggestion can however be manually adjusted by the user. In Fig. \ref{fig:GUI_fstar} we display a screenshot of the selection of the Nyqvist frequency.

\subsection{Command-Line Interface}
The command-line interface is designed to easily embed the complete analysis in scripts. All the parameters have to be specified as command line arguments (see Tab.~\ref{tab:CLI}). The program outputs a PDF file with several plots, and textual or binary files containing the raw results of the calculation. Those files can be easily used for further analysis.
\begin{table}[htb]
    \centering
    \begin{small}
    \begin{tabular}{l|l}
          \Verb|--input-format {table,dict,lammps}| & input format \\
          \Verb|-k MAINFLUXKEY| & the header of the $J_0$ flux column \\ 
          \Verb|-j ADD_CURRENTS| & header of additional current (optional) \\
          \Verb|-C {electric,heat,stress}| & the type of current \\
          \Verb|-u {real,metal,qepw,GPa...}| & units \\
          \Verb|-t TIMESTEP| & timestep in fs \\
          \Verb|--VOLUME VOLUME| & volume of the system in $\angstrom^3$ \\
          \Verb|--TEMPERATURE TEMPERATURE| & temperature of the system in K \\
          \Verb|-r| & resample the time series with the specified FSTAR \\
          \Verb|--FSTAR FSTAR| & maximum frequency to analyse in THz \\
          \Verb|-w PSD_FILTERW| & size of the moving average filter used in the plots, \\ &\quad for visualisation purposes (THz) \\
          \Verb|--help| & show the complete help \\
          \Verb|--list-currents| & list all the currents and units implemented in the code\\
          \Verb|INPUTFILE| & input file
    \end{tabular}
    \end{small}
    \caption{List of important command line parameters.}
    \label{tab:CLI}
\end{table}

\subsection{Python API description}
Here we provide an example of the usage of the code as a Python library. In this example, that is also provided in the git repository \cite{sportran} and documentation, we perform the analysis of the energy current time series of a molten salt, sodium chloride. We stress that the procedure is the same for any type of transport coefficient calculation. The data is contained in a plain text file that was extracted from the output of a LAMMPS \cite{PLIMPTON19951} simulation, formatted as follows:
\begin{small}
\begin{Verbatim}[fontsize=\small, frame=lines]
 Temp  c_flux[1] c_flux[2] c_flux[3] c_vcm[1][1] c_vcm[1][2] c_vcm[1][3]
 1442.7319  250.86549 20.619423 200.115 -0.15991832 -0.071370426 0.020687917
 1440.8060  196.22265 82.667342 284.3325 -0.13755206 -0.071002931 -0.011279876
 ...
\end{Verbatim}
\end{small}
Note the first line and the LAMMPS-like notation to define vector components. Besides the fluxes, there can be additional columns with scalar data, \emph{e.g.} the temperature. In this particular case, the first vector quantity named \verb.c_flux. is the energy current, while the second vector quantity named \Verb.c_vcm[1]. is the velocity of the center of mass of the sodium atoms. This file contains time series that can be generated by any MD code.
The API provides a generic module to read this kind of data files:
\begin{small}
\begin{Verbatim}[fontsize=\small, frame=lines]
 import sportran as st
 jfile = st.i_o.TableFile('./examples/data/NaCl.dat', group_vectors=True)
 jfile.read_datalines(
        start_step=0, NSTEPS=0, select_ckeys=['Temp', 'flux', 'vcm[1]'])
\end{Verbatim}
\end{small}
The \Verb|select_ckeys| argument must be set with the names of the desired columns. The code will automatically read them as Cartesian components of a vector-valued process.
Please refer to the documentation for additional input formats.

We are now ready to initialise a \texttt{Current}-type object. \texttt{Current} is an abstract class that serves as a template for different types of currents. The \texttt{GenericCurrent} class, derived from \texttt{Current}, defines a generic current time series: in order to estimate the transport coefficient, only the time step and the $\kappa_{\mathrm{scale}}$ factor must be defined. To streamline the user workflow, a few other specialised subclasses are available: \texttt{HeatCurrent} (thermal conductivity), \texttt{ElectrictCurrent} (electrical conductivity), \texttt{StressCurrent} (viscosity), each corresponding to different types of transport processes. For each current type several units are available,\footnote{A list of units for each current subclass can be printed: \emph{e.g.} \Verb|print(sp.current.HeatCurrent.get_units_list())|.} that simply define the $\kappa_{\mathrm{scale}}$ factor. In the case of this example:
\begin{small}
\begin{Verbatim}[fontsize=\small, frame=lines]
 DT_FS = 5.0            # time step [fs]
 TEMPERATURE = np.mean(jfile.data['Temp'])  # mean temperature [K]
 VOLUME = 40.21**3      # volume [A^3]
 j = st.HeatCurrent([jfile.data['flux'], jfile.data['vcm[1]']],
         UNITS='metal', DT_FS=DT_FS, TEMPERATURE=TEMPERATURE, VOLUME=VOLUME)
\end{Verbatim}
\end{small}
\Verb|metal| are the units used to compute the heat flux (as defined in LAMMPS \cite{PLIMPTON19951}). The input parameters needed to define each \Verb|Current| subclass are listed in the code documentation.

\begin{figure}[tb]
    \centering
    \includegraphics[width=\textwidth]{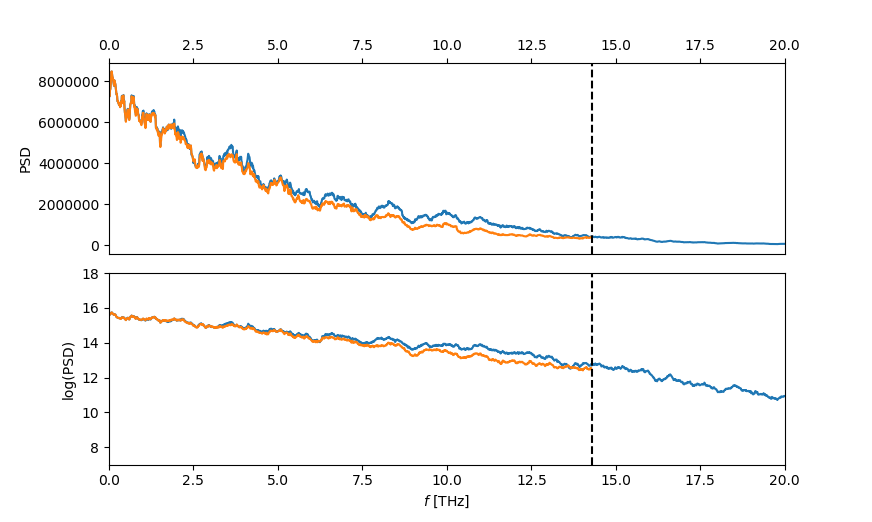}
    \caption{Resampled time series. The blue and orange lines represent $\bar{\mathcal{S}}_k^0$ of the original and the resampled time series, respectively.}
    \label{fig:psd_plot_resampled}
\end{figure}

We are now in the position to compute the sample $\bar{\mathcal{S}}_k^0$. Remember again that in the univariate case this is simply the periodogram of the original time series. In order to plot $\bar{\mathcal{S}}_k^0$ and its logarithm we can use the following function:
\begin{Verbatim}[fontsize=\small, frame=lines]
 ax = j.plot_periodogram(PSD_FILTER_W=0.4, kappa_units=True,
                         label=r'$\bar{\mathcal{S}}^0_k$')
\end{Verbatim}
\begin{figure}[tb]
    \centering
    \includegraphics[width=\textwidth]{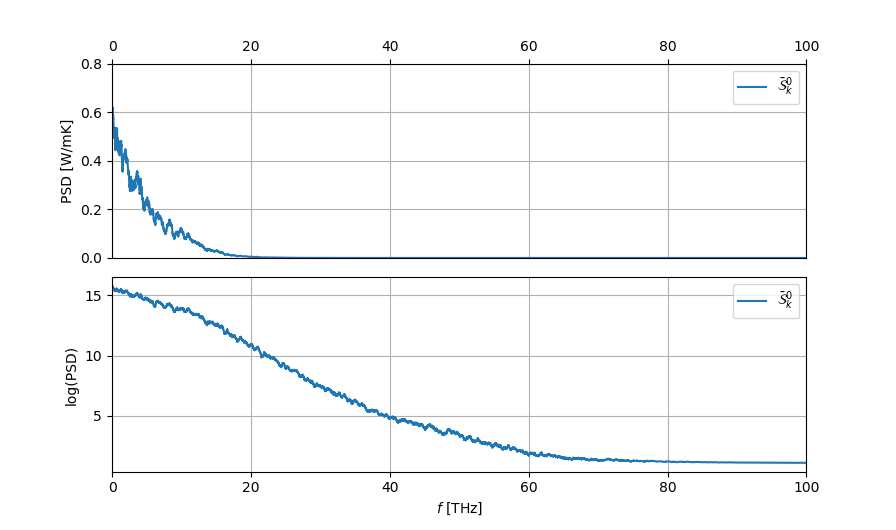}
    \caption{Plot of $\bar{\mathcal{S}}_k^0$ and $\log \bar{\mathcal{S}}_k^0$ generated using the class method \texttt{HeatCurrent.plot\_periodogram()}.}
    \label{fig:psd_plot}
\end{figure}
\Verb|PSD_FILTER_W| defines the width in THz of a moving average filter used for visualisation purposes. The result is shown in Fig.~\ref{fig:psd_plot}. Since we are interested in the zero-frequency value of $\bar{\mathcal{S}}_k^0$, we resample the time series in order to decrease the Nyqvist frequency to $f^*$ and focus on the lower part of the spectrum. We do this as follows:
\begin{center}
\begin{Verbatim}[fontsize=\small, frame=lines]
 FSTAR_THZ = 14.0
 jf, ax = j.resample(fstar_THz=FSTAR_THZ, plot=True, freq_units='thz')
 ax.set_xlim([0, 20])
\end{Verbatim}
\end{center}
The resulting plot is shown in Fig.~\ref{fig:psd_plot_resampled}. 
We are now ready to perform the cepstral analysis. This is as simple as calling the function:
\begin{center}
\begin{Verbatim}[fontsize=\small, frame=lines]
 jf.cepstral_analysis()
\end{Verbatim}
\end{center}
The result provided is the following:
\begin{center}
\begin{Verbatim}[fontsize=\small, frame=lines]
 -----------------------------------------------------
 CEPSTRAL ANALYSIS
 -----------------------------------------------------
 AIC_Kmin  = 3  (P* = 4, corr_factor = 1.000000)
 L_0*   =          15.158757 +/-   0.056227
 S_0*   =     6824108.702608 +/- 383697.095268
 -----------------------------------------------------
 kappa* =           0.498310 +/-   0.028018  W/mK
 -----------------------------------------------------
\end{Verbatim}
\end{center}
For additional details please refer to the Jupyter notebook examples provided with the source code, or to the documentation.

All the steps and plots of this workflow are implemented in the command line tool and in the graphical user interface, that streamlines the execution for the end user.

\subsection{Extending \SporTran\ and code details}

We remark again that the core part of the code is process-agnostic, \emph{i.e.} it does not depend on the type of transport coefficient, and can be seen as a way to compute \emph{any} Green-Kubo integral from a flux time series. 
However, when performing an actual calculation, the ability to set the physical units can be very practical. 
For this reason, \SporTran\  is designed in a way that makes it easy to add new custom units and transport coefficients that can be expressed as in Eq.~\eqref{eq:multi-kappa2}, where physical coefficients such as temperature and volume are multiplied. This enables the user to input all the required parameters in a simple and friendly manner through all the available interfaces, and to get back the result in the preferred units. If the following procedure is used, all the built-in interfaces (GUI and CLI) will detect the new parameters and request them when needed, thanks to the introspection features of the Python language.

Adding a new unit is very easy: it is sufficient to go to the folder \Verb|sportran/current/units|, open the module corresponding to the desired current type (heat, electric, stress, ...) and add a function named \Verb|kappa_scale_mynewunitname|, similarly to the other functions found therein. 
The code will list this new unit called \Verb|mynewunitname| in every user interface  and in the help function.

In order to define a new type of current and set its units, the following operations are required.
First, in the folder \Verb|sportran/current| add a new module (\emph{e.g.} \Verb|mycurrent.py|), in which a subclass of \Verb|sportran.current.Current| (\emph{e.g.} \Verb|MyCurrent|) is defined. 
Then:
\begin{itemize}
\item Define the following class attributes:
    \begin{itemize}
    \item \Verb|_current_type|: the name of the current. This will be used by the user interfaces and to look for the available units corresponding to this current (\emph{e.g.} \Verb|mycurrent|).
    \item \Verb|_input_parameters|: a set of parameters that this class needs, \emph{e.g. } \Verb|{'DT_FS', 'UNITS', 'TEMPERATURE', 'VOLUME'}|.
    \item \Verb|_KAPPA_SI_UNITS|: a string describing the units of the transport coefficient, \emph{e.g.} \Verb|W/m/K|.
    \end{itemize}
\item Define a \Verb|_builder| method (property). This is a method returning a dictionary of all the parameters needed to rebuild an identical current object.
\item Define the units: create a new module in the folder \Verb|sportran/current/units| and name it after \Verb|_current_type| (\emph{e.g.} \Verb|mycurrent.py|). Add here all the desired units as functions called \Verb|kappa_scale_myunit|, as previously explained. 
The input parameters of these functions must be the same listed in the \Verb|_input_parameters| class attribute.
\end{itemize}

The code has an extensive documentation written with Sphinx available at \url{https://sportran.readthedocs.io} and a complete test suite that the user is strongly advised to run before using the code on a new machine, or after modifying the code. 
Examples are included in the package, both for the single-component and the multi-component case in the form of Jupyter notebooks and command-line scripts. Any issue reports or contributions to this code are encouraged and can be submitted to the GitHub page \cite{sportran}.

\section*{Declaration of competing interests}
The authors declare that they have no known competing financial interests or personal relationships that could have appeared to influence the work reported in this paper.

\section*{Acknowledgements}
The authors are grateful to Federico Grasselli, Aris Marcolongo, Paolo Pegolo, and Davide Tisi for enlightening discussions throughout the development of the code and the writing of this paper. This work was partially funded by the EU through the MaX Centre of Excellence for supercomputing applications (Project No. 824143) and by the Italian Ministry for University and Research (MUR), through the PRIN grant FERMAT. Sebastiano Bisacchi wishes to thank the Quantum ESPRESSO Foundation for funding a Summer internship for high-school students at SISSA, where the GUI was developed in the Summer of 2019.





\bibliographystyle{elsarticle-num-notitle}
\bibliography{bib.bib}

\begin{thebibliography}{10}
\expandafter\ifx\csname url\endcsname\relax
  \def\url#1{\texttt{#1}}\fi
\expandafter\ifx\csname urlprefix\endcsname\relax\def\urlprefix{URL }\fi
\expandafter\ifx\csname href\endcsname\relax
  \def\href#1#2{#2} \def\path#1{#1}\fi

\bibitem{Ercole2017}
L.~Ercole, A.~Marcolongo, S.~Baroni, Sci. Rep. 7 (2017) 15835.
\newblock \href {https://doi.org/10.1038/s41598-017-15843-2}
  {\path{doi:10.1038/s41598-017-15843-2}}.

\bibitem{Bertossa2019}
R.~Bertossa, F.~Grasselli, L.~Ercole, S.~Baroni, Phys. Rev. Lett. 122 (2019)
  255901.
\newblock \href {https://doi.org/10.1103/PhysRevLett.122.255901}
  {\path{doi:10.1103/PhysRevLett.122.255901}}.

\bibitem{Marcolongo2016}
A.~Marcolongo, P.~Umari, S.~Baroni, Nature Phys. 12 (2016) 80--84.
\newblock \href {https://doi.org/10.1038/nphys3509}
  {\path{doi:10.1038/nphys3509}}.

\bibitem{qeheat}
A.~Marcolongo, R.~Bertossa, D.~Tisi, S.~Baroni,
  \href{https://www.sciencedirect.com/science/article/pii/S0010465521002022}{Computer
  Physics Communications} 269 (2021) 108090.
\newblock \href {https://doi.org/https://doi.org/10.1016/j.cpc.2021.108090}
  {\path{doi:https://doi.org/10.1016/j.cpc.2021.108090}}.
\newline\urlprefix\url{https://www.sciencedirect.com/science/article/pii/S0010465521002022}

\bibitem{Marcolongo2020}
A.~Marcolongo, L.~Ercole, S.~Baroni,
  \href{https://doi.org/10.1021/acs.jctc.9b01174}{Journal of Chemical Theory
  and Computation} 16~(5) (2020) 3352--3362.
\newblock \href {http://arxiv.org/abs/https://doi.org/10.1021/acs.jctc.9b01174}
  {\path{arXiv:https://doi.org/10.1021/acs.jctc.9b01174}}, \href
  {https://doi.org/10.1021/acs.jctc.9b01174}
  {\path{doi:10.1021/acs.jctc.9b01174}}.
\newline\urlprefix\url{https://doi.org/10.1021/acs.jctc.9b01174}

\bibitem{Grasselli2019}
F.~Grasselli, S.~Baroni, Nature Physics 15 (2019) 809--813.
\newblock \href {https://doi.org/10.1038/s41567-019-0562-0}
  {\path{doi:10.1038/s41567-019-0562-0}}.

\bibitem{Tisi2021}
D.~Tisi, L.~Zhang, R.~Bertossa, H.~Wang, R.~Car, S.~Baroni,
  \href{https://link.aps.org/doi/10.1103/PhysRevB.104.224202}{Phys. Rev. B} 104
  (2021) 224202.
\newblock \href {https://doi.org/10.1103/PhysRevB.104.224202}
  {\path{doi:10.1103/PhysRevB.104.224202}}.
\newline\urlprefix\url{https://link.aps.org/doi/10.1103/PhysRevB.104.224202}

\bibitem{Bogert1963}
B.~P. Bogert, J.~R. Healy, J.~W. Tukey, in: Proceedings of the Symposium on
  Time Series Analysis, 1963, pp. 209--243.

\bibitem{Oppenheim:2004}
A.~Oppenheim, R.~Schafer,
  \href{http://ieeexplore.ieee.org/document/1328092/}{IEEE Signal Processing
  Magazine} 21~(5) (2004) 95--106.
\newblock \href {https://doi.org/10.1109/MSP.2004.1328092}
  {\path{doi:10.1109/MSP.2004.1328092}}.
\newline\urlprefix\url{http://ieeexplore.ieee.org/document/1328092/}

\bibitem{Baroni2018}
S.~Baroni, R.~Bertossa, L.~Ercole, F.~Grasselli, A.~Marcolongo, Heat Transport
  in Insulators from Ab Initio Green-Kubo Theory, 2nd Edition, Springer
  International Publishing, Cham, 2018, pp. 1--36.
\newblock \href {http://arxiv.org/abs/1802.08006} {\path{arXiv:1802.08006}},
  \href {https://doi.org/10.1007/978-3-319-50257-1\_12-1}
  {\path{doi:10.1007/978-3-319-50257-1\_12-1}}.

\bibitem{harris2020array}
C.~R. Harris, K.~J. Millman, S.~J. van~der Walt, R.~Gommers, P.~Virtanen,
  D.~Cournapeau, E.~Wieser, J.~Taylor, S.~Berg, N.~J. Smith, R.~Kern, M.~Picus,
  S.~Hoyer, M.~H. van Kerkwijk, M.~Brett, A.~Haldane, J.~F. del R{\'{i}}o,
  M.~Wiebe, P.~Peterson, P.~G{\'{e}}rard-Marchant, K.~Sheppard, T.~Reddy,
  W.~Weckesser, H.~Abbasi, C.~Gohlke, T.~E. Oliphant, Nature 585~(7825) (2020)
  357--362.
\newblock \href {https://doi.org/10.1038/s41586-020-2649-2}
  {\path{doi:10.1038/s41586-020-2649-2}}.

\bibitem{sportran}
L.~Ercole, R.~Bertossa, \url{https://github.com/sissaschool/sportran}
  (2017-2022). The \texttt{SporTran} package was previously known as
  ``\texttt{thermocepstrum}'' and can be accessed at URL
  \url{https://github.com/lorisercole/thermocepstrum} and
  \url{https://github.com/lorisercole/sportran} as well. Documentation
  available at \url{https://sportran.readthedocs.io}.

\bibitem{Onsager1931a}
L.~Onsager, Phys. Rev. 37~(4) (1931) 405--426.
\newblock \href {https://doi.org/10.1103/PhysRev.37.405}
  {\path{doi:10.1103/PhysRev.37.405}}.

\bibitem{Onsager1931b}
L.~Onsager, Phys. Rev. 38 (1931) 2265.
\newblock \href {https://doi.org/10.1103/PhysRev.38.2265}
  {\path{doi:10.1103/PhysRev.38.2265}}.

\bibitem{Green1952}
M.~S. Green, J. Chem. Phys. 20~(8) (1952) 1281--1295.
\newblock \href {https://doi.org/10.1063/1.1700722}
  {\path{doi:10.1063/1.1700722}}.

\bibitem{Green1954}
M.~S. Green, J. Chem. Phys. 22 (1954) 398--413.
\newblock \href {https://doi.org/10.1063/1.1740082}
  {\path{doi:10.1063/1.1740082}}.

\bibitem{Kubo1957a}
R.~Kubo, J. Phys. Soc. Jpn. 12~(6) (1957) 570--586.
\newblock \href {https://doi.org/10.1143/JPSJ.12.570}
  {\path{doi:10.1143/JPSJ.12.570}}.

\bibitem{Kubo1957b}
R.~Kubo, M.~Yokota, S.~Nakajima, J. Phys. Soc. Jpn. 12~(11) (1957) 1203--1211.
\newblock \href {https://doi.org/10.1143/JPSJ.12.1203}
  {\path{doi:10.1143/JPSJ.12.1203}}.

\bibitem{Ercole2016}
L.~Ercole, A.~Marcolongo, P.~Umari, S.~Baroni, J. Low Temp. Phys. 185 (2016)
  79--86.
\newblock \href {https://doi.org/10.1007/s10909-016-1617-6}
  {\path{doi:10.1007/s10909-016-1617-6}}.

\bibitem{Akaike1973}
H.~Akaike, Information theory and an extension of the maximum likelihood
  principle, in 2nd International Symposium on Information Theory, edited by B.
  N. Petrov and F. Csáki, 1972.

\bibitem{Akaike1974}
H.~Akaike, \href{https://doi.org/10.1109/TAC.1974.1100705}{IEEE Trans. Autom.
  Control} 19~(6) (1974) 716--723.
\newblock \href {https://doi.org/10.1109/TAC.1974.1100705}
  {\path{doi:10.1109/TAC.1974.1100705}}.
\newline\urlprefix\url{https://doi.org/10.1109/TAC.1974.1100705}

\bibitem{PolyGamma}
E.~W. Weisstein,
  \href{http://mathworld.wolfram.com/PolygammaFunction.html}{from MathWorld --
  A Wolfram Web Resource
  \url{http://mathworld.wolfram.com/PolygammaFunction.html}}.
\newline\urlprefix\url{http://mathworld.wolfram.com/PolygammaFunction.html}

\bibitem{PLIMPTON19951}
S.~Plimpton, J. Comput. Phys. 117~(1) (1995) 1 -- 19.
\newblock \href {https://doi.org/10.1006/jcph.1995.1039}
  {\path{doi:10.1006/jcph.1995.1039}}.

\end{thebibliography}


\begin{thebibliography}{0}
\bibitem{1} L. Ercole, A. Marcolongo, and S. Baroni, Sci. Rep. 7, 15835 (2017);
\bibitem{2} R. Bertossa, F. Grasselli, L. Ercole, and S. Baroni Phys. Rev. Lett. 122, 255901 (2019).
\bibitem{3} S. Baroni, R. Bertossa, L. Ercole, F. Grasselli, and A. Marcolongo, in \emph{Handbook of Materials Modeling. Applications: Current and Emerging Materials}, edited by W. Andreoni and S. Yip (Springer, 2018) 2nd ed., Chap. 12-1 (\url{https://arxiv.org/abs/1802.08006});
\end{thebibliography}

\end{document}